\documentclass[11pt,a4paper]{scrartcl}

\usepackage{latexsym}
\usepackage{amssymb}
\usepackage{amsmath}
\usepackage[authoryear, round]{natbib}

\usepackage{ifpdf}

\ifpdf
    \usepackage[pdftex]{graphicx}
    \usepackage{color}
    \definecolor{myred}{rgb}{0.5,0,0}
    \definecolor{myblue}{rgb}{0,0,0.75}
    \definecolor{mygreen}{rgb}{0,0.5,0}
   \usepackage[pdftex,
               pdftitle={The two defaults scenario for stressing credit portfolio loss distributions},
               pdfsubject={Risk concentration, stress test, scenario analysis, joint default probability},
               pdfkeywords={},
               pdfauthor={Dirk Tasche},
               pdfstartview=FitR,
               breaklinks=true,
               colorlinks=true,
               citecolor=myred,
               linkcolor=myblue,
               urlcolor=mygreen]{hyperref}
\else
    \usepackage[dvips]{graphicx}
   \usepackage{hyperref}
\fi

\newtheorem{theorem}{Theorem}[section]

\newtheorem{remark}[theorem]{Remark}
\newtheorem{example}[theorem]{Example}
\newtheorem{proposition}[theorem]{Proposition}
\newtheorem{definition}[theorem]{Definition}
\newtheorem{corollary}[theorem]{Corollary}

\numberwithin{equation}{section}

\begin{document}
\title{The two defaults scenario for stressing credit portfolio loss distributions}

\author{%
Dirk Tasche\thanks{E-mail: dirk.tasche@gmx.net\newline
The author currently works at the Prudential Regulation Authority (a
directorate of the Bank of England). He is also a visiting professor at
Imperial College, London. 
The opinions expressed in this paper are those of the author 
and do not necessarily reflect views of the Bank of England.}}

\date{}
\maketitle

\begin{abstract}
The impact of a stress scenario of default events on the loss distribution of
a credit portfolio can be assessed by determining the loss distribution conditional on these events. 
While it is conceptually easy to estimate 
loss distributions conditional on default events by means of Monte Carlo
simulation, it becomes impractical for two or more simultaneous defaults
as then the conditioning event is extremely rare. 
We provide an analytical approach to the calculation of the conditional loss distribution for the
CreditRisk$^+$ portfolio model with independent random loss given default
distributions. 
The analytical solution for this case can be used to check the accuracy of an approximation to
the conditional loss distribution whereby the unconditional model is run with stressed input probabilities of default (PDs).
It turns out that this approximation is unbiased. Numerical examples, however, suggest that 
the approximation may be seriously inaccurate but that the inaccuracy leads to overestimation of 
tail losses and hence the approach errs on the conservative side.
\end{abstract}



\section{Introduction}\label{tas_sec_0}

Stress test scenarios for credit risk typically are stated in terms of economic factors but sometimes involve 
defaults of larger counterparties or obligors \citep[see e.g.][Section~10.3.3]{StressGuidance2015}.
Default of a large obligor not only has a direct impact on the 
profit and loss of a bank and potentially also on its capital basis. Due to mutual 
dependence of default events the default of one or more obligors can have a significant
impact on the loss distribution of the remaining portfolio, too. Determination of credit portfolio loss
distributions conditional on defaults, therefore, can be considered a special stress testing
technique. Such analysis, in particular, can help to decide whether a large exposure to a certain
obligor is just a risk concentration because of its size or, even worse, also significant
part of a sector or industry risk concentration.
Loss distributions conditional on default of one or more obligors therefore are promising means to identify 
vulnerabilities of banks.

Techniques for measuring the impact of macro-economic stress scenarios on credit portfolio losses are 
well-established \citep[see e.g.][]{Bonti&Kalkbrener&Lotz&Stahl, kalkbrener2015stress}. In particular,
it is common and efficient to analyse such stress scenarios by means of Monte-Carlo simulation.
In principle, it is easy to determine also the impact of the default of one or more obligors 
via a Monte-Carlo simulation approach: Just eliminate all simulation iterations from 
the sample in which the obligor(s) on whose default(s) conditioning is to be conducted have not 
defaulted. This is feasible in practice for one default, but becomes impracticable for two or more
defaults.

An obvious approach to try and work around this problem would be to deploy an unstressed (i.e.\ unconditional) model
for the analysis but to feed it with parameters like probabilities of default (PDs) and loss-given-default
(LGD) that have been stressed in a separate exercise before. This approach -- which may be called
`stressed input parameters' approach --, however, might fail to
fully capture the dependence structure of the model and its changes under stress such that misjudgement
of the stress impact could be the consequence.

This paper seeks to assess how accurate the results calculated with the `stressed input parameters' approach
are when compared to results from a fully-fledged conditional loss distribution approach. For that purpose we
revisit the CreditRisk$^+$ credit portfolio risk model \citep{CRplus} and derive a 
representation of the loss distribution conditional on the default of two fixed obligors that allows for
the computation of the distribution without Monte-Carlo simulation. Two numerical examples then suggest
that results from the `stressed input parameters' approach may be seriously inaccurate but tend to be inaccurate
in a conservative direction and to overestimate tail losses. 

\citet[][equation (3.31)]{Tasche2004} showed how the loss distribution conditional on one 
default can be calculated analytically in the CreditRisk$^+$ model with random loss severities.
In this paper the related formulas for the case of two defaults are provided. Formulas for the
cases of three or more defaults can be readily derived in the same way as the formula for
the case of two defaults is derived. As a consequence of the likely lack of practical relevance
of cases of three or more defaults scenarios, we do not provide the results for these cases here. 
Moreover, the paper is focused on the theoretical derivation of the main result on the loss distribution 
conditional on two defaults and its interpretation. The question
of practical numerical implementation is only considered to such an extent as needed for 
the numerical examples.

The plan of this paper is as follows:
\begin{itemize}
	\item As background and for introducing the notation, 
	Section~\ref{tas_sec_3} provides a description of the CreditRisk$^+$ model as presented in \citet{CRplus} or
	\citet{Gundlach2004}. The CreditRisk$^+$ model described here is enhanced to allow for random loss severities\footnote{\citet{Schmock2008} describes
	a further generalisation of the model to include connected groups of obligors.}.
	\item In Section \ref{tas_sec_4} the results on the conditional loss distributions are presented 
	and their application is discussed. 
	To derive the results we revisit the approach used in \citet{Tasche2004} to develop analytical representations
of the Value-at-Risk and Expected Shortfall contributions of single obligors in CreditRisk$^+$.
    \item In Section~\ref{sec:StressedParameters} we present the technical particulars of the
    `stressed input parameters' approach and prove that the first moments of the resulting loss
    distributions are the same as the first moments of the proper loss distribution conditional on two 
    defaults.
    \item Section~\ref{sec:Examples} provides two numerical examples to shed light on the question
    of how close the results from the `stressed input parameters' and conditional 
    loss distribution approaches are in general.
	\item The paper concludes with summarising comments in Section \ref{sec:concl}.
\end{itemize}

\section{An analytical credit portfolio model with random loss severities}\label{tas_sec_3}

The approach to the CreditRisk$^+$
loss distribution as described in \citet{CRplus} or
	\citet{Gundlach2004} is driven by analytical
considerations and -- to some extent -- hides the way in which the Poisson
approximation is used to smooth the loss distribution. While preserving the
notation of \citet{Gundlach2004}, therefore we review in this
section the steps that lead to the formula for the generating function 
of the loss distribution in
\citet{CRplus} and \citet{Gundlach2004}. When doing so, we 
slightly generalize the methodology to the case of stochastic exposures -- thus allowing for 
random loss severities -- that
are independent of the default events and the random factors expressing the
dependence on sectors or industries. This generalization can be afforded at no
extra cost as the result is again a generating function in the shape as 
presented in \citet[][equation (2.19)]{Gundlach2004}, the only difference being
that the sector polynomials are composed another way.

Write $\mathbf{1}_A$ for the default indicator of obligor $A$, i.e.
$\mathbf{1}_A = 0$ if $A$ does not default in the observation period and 
$\mathbf{1}_A = 1$ if $A$ defaults.
In \citet{CRplus} and \citet{Gundlach2004}, an approximation is derived for the
distribution of the portfolio loss variable $X = \sum_A \mathbf{1}_A\,\nu_A$
with the  $\nu_A$ denoting deterministic potential losses. A careful
inspection of the beginning of Section~5 of
\citet{Gundlach2004} reveals that the main step in the approximation
procedure is to replace the $\{0, 1\}$-valued indicators $\mathbf{1}_A$ 
by integer-valued random variables
$D_A$ with the same expected values. These variables $D_A$ are conditionally
Poisson distributed given some economic factors $S_1, \ldots, S_N$.

Here, we want to study the distribution of the more general loss variable $X =
\sum_A \mathbf{1}_A\,\mathcal{E}_A$, where $\mathcal{E}_A$ denotes the random
outstanding exposure of obligor $A$. We assume that $\mathcal{E}_A$ takes on
non-negative integer values. However, just replacing $\mathbf{1}_A$ by $D_A$ as in
the case of deterministic potential losses does not yield a nice generating
function -- ``nice'' in the sense that the CreditRisk$^+$ algorithms for
extracting the loss distribution can be applied. We instead consider the
approximate loss variable
\begin{subequations}
\begin{equation}
  \label{tas_eq_loss}
  X\,=\,\sum_A \sum_{i=1}^{D_A} \mathcal{E}_{A, i},
\end{equation}
where $\mathcal{E}_{A, 1}, \mathcal{E}_{A, 2}, \ldots$ are independent copies
of $\mathcal{E}_A$. Thus, we approximate the terms
$\mathbf{1}_A\,\mathcal{E}_A$ by conditionally compound Poisson sums. For the
sake of brevity, we write
\begin{equation}\label{tas_eq_ind_loss}
  Y_A\,=\,\sum_{i=1}^{D_A} \mathcal{E}_{A, i}
\end{equation}
for the loss suffered due to obligor $A$. A careful inspection of the
arguments presented to derive Equation~(2.19) of \citet{Gundlach2004} now yields the following result on the
generating function of the distribution of the loss variable $X$.
\end{subequations}
\begin{subequations}
\begin{theorem}
  \label{tas_th_3}
Define the ``loss'' variable $X$ by
(\ref{tas_eq_loss}) and specify the distribution of $X$ by the following assumptions:
\begin{itemize}
\item[(i)] The approximate default indicators $D_A$ are conditionally independent given a vector of ``economic'' factors 
	$\mathbf{S} = (S_0, S_1,
  \ldots, S_N)$. The conditional distribution of $D_A$ given $\mathbf{S}$ is Poisson
  with intensity $p_A^S \,=\,p_A \sum_{k=0}^N w_{A k}\,S_k$ where $p_A > 0$ denotes 
  the ``probability of default'' (PD) of obligor $A$ and $0 \le w_{A k} \le 1$ are 
  ``factor loadings'' such that $\sum_{k=0}^N w_{A k} = 1$ for each obligor $A$.
\item[(ii)] The \emph{idiosyncratic} factor $S_0$ is a constant and equals 1. The factors $S_1,
  \ldots, S_N$ are independent and Gamma-distributed\footnote{%
  We call a positive random variable $Y$ Gamma-distributed if it has a density
  $f_Y(y) = \frac{y^{\alpha-1}}{\beta^\alpha\,\Gamma(\alpha)}\,e^{-y/\beta}$, $y >0$ for some parameters
  $\alpha >0$, $\beta>0$. The function $\Gamma$ denotes the familiar Gamma function generalising the factorial.} 
  with unit expectations $\mathbb{E}[S_k] = 1$  
  and parameters\footnote{%
  $\beta_k = 1/\alpha_k$ is implied by the assumption that $S_k$ has unit expectation.} 
  $(\alpha_k, \beta_k) = (\alpha_k,
  1/\alpha_k)$ for $k=1,\ldots, N$.
\item[(iii)] The random variables $\mathcal{E}_{A, 1}, \mathcal{E}_{A, 2}, \ldots$
  are independent copies of a non-negative integer-valued random variable
  $\mathcal{E}_A$ and, additionally, are also independent of the $D_A$ and $\mathbf{S}$.
The distribution of 
$\mathcal{E}_A$ is given by its generating function
\begin{equation}
  \label{tas_eq_exposure}
  H_A(z) \,=\,\mathbb{E}\bigl[z^{\mathcal{E}_A}\bigr], \quad z \in \mathbb{C}, |z| \le 1.
\end{equation}
\end{itemize}
Define for $k = 0,1, \ldots, N$ the \emph{sector
polynomial} $\mathcal{Q}_k$ by
\begin{equation}
  \label{tas_eq_polynomial}
\mathcal{Q}_k(z)\,=\,\frac{1}{\mu_k} \sum_A w_{Ak}\,p_A\,H_A(z),
\end{equation}
where the \emph{sector default intensities} $\mu_k$ are given by 
\begin{equation}
	\mu_k \ = \ \sum_A w_{Ak}\,p_A.
\end{equation}

Then the generating function $G_X(z) = \mathbb{E}[z^X]$, $z \in \mathbb{C}, |z| \le 1$, 
of the loss variable $X$ can be represented
as
\begin{equation}
  \label{tas_eq_gen}
  G_X(z)\,=\,e^{\mu_0\,(\mathcal{Q}_0(z)-1)} \prod_{k=1}^N
  \left(\frac{1-\delta_k}
{1-\delta_k\,\mathcal{Q}_k(z)}\right)^{\alpha_k},
\end{equation}
where the constants $\delta_k$ are defined as $\delta_k = \mu_k / (\mu_k + \alpha_k)$.
\end{theorem}
\end{subequations}
\begin{remark}\
  \label{tas_rm_0}
  \begin{enumerate}
  \item[(i)] The case of deterministic severities can be regained from
  Theorem~\ref{tas_th_3} by choosing the exposures constant, e.g.\
$\mathcal{E}_A = \nu_A$. Then the generating functions of the exposures are
  just monomials, namely $H_A(z) = z^{\nu_A}$.
\item[(ii)] Representation \eqref{tas_eq_gen} of the generating function of the
portfolio loss distribution implies that the portfolio loss distribution can
be interpreted as the distribution of a sum of $N+1$ independent \emph{sector
loss distributions} that correspond to the economic factors $(S_0, S_1, \ldots, S_N)$.\\
The term $e^{\mu_0\,(\mathcal{Q}_0(z)-1)}$ is the generating function of a random variable with
a compound\footnote{See, e.g., \citet{Rolski&al} for background information on compound 
distributions and generating functions.} Poisson distribution that can be realised as $\sum_{i=1}^{T_0} \eta_{0,i}$ where
$T_0, \eta_{0,1}, \eta_{0,2}, \ldots$ are independent, $T_0$ is Poisson-distributed with
intensity $\mu_0$, and $\eta_{0,1}, \eta_{0,2}, \ldots$ are i.i.d.\ with generating function
$\mathcal{Q}_0(z)$.\\
The terms $\left(\frac{1-\delta_k}{1-\delta_k\,\mathcal{Q}_k(z)}\right)^{\alpha_k}$, $k = 1, \ldots, N$,
are the generating functions of random variables with compound negative binomial distributions that
can be realised as $\sum_{i=1}^{T_k} \eta_{k,i}$ where
$T_k, \eta_{k,1}, \eta_{k,2}, \ldots$ are independent, $T_k$ is negative binomially distributed\footnote{%
We call a random variable $Y$ with values in the non-negative integers negative binomially 
distributed with size parameter $a>0$ and failure probability $0 < p < 1$ if 
$\mathbb{P}[Y=k] = \frac{\Gamma(a+k)}{\Gamma(a)\,k!}\,(1-p)^a\,p^k$ for $k=0,1,2, \ldots$.
If the size parameter $a$ of a negative binomial distribution is a positive integer then
the distribution can be interpreted as the distribution of the number of failures in a series of
independent identical experiments before the $a$-th success is observed.} with
\emph{failure probability} $\delta_k$ and \emph{size parameter} $\alpha_k$, 
and $\eta_{k,1}, \eta_{k,2}, \ldots$ are i.i.d.\ with generating function
$\mathcal{Q}_k(z)$.\\
With this representation of the portfolio loss distribution as the convolution of compound Poisson and
negative binomial distributions, the sector polynomials $\mathcal{Q}_k$ can be interpreted as 
the generating functions of typical loss severities in the respective sectors.
  \end{enumerate}
\end{remark}
By means of Theorem \ref{tas_th_3} the loss distribution of the generalized
model (\ref{tas_eq_loss}) can be calculated in principle with the same algorithms as in the
case of the original CreditRisk$^+$ model. Once the probabilities
$\mathbb{P}[X=x]$, $x$ non-negative integer, are known, it is an easy task to
calculate the loss quantiles $q_\theta(X)$ as defined by 
\begin{equation}\label{tas_eq_3}
q_\theta(X) \ = \ \min\{x \ge 0: \mathrm{P}[X \le x]	\ge \theta\},
\end{equation}
or related risk measures like Value-at-Risk or Expected Shortfall.

When working with Theorem~\ref{tas_th_3}, one has to decide whether random
exposures shall be taken into account, and in case of a decision in favour of
doing so, how the exposure distributions are to be modeled. \citet[][Example~1]{Tasche2004}
and \citet{Schmock2008} present some possible choices of discrete exposure distributions.
\citet{Gordy2004} discusses an approximate but similar approach to random severities 
with continuous distributions.

Parametrisation of the factor model as described in Theorem~\ref{tas_th_3} is non-trivial because
the assumption of independent economic factors is unrealistic in practice. 
\citet{vandendorpe2008parameterization} discuss how to derive appropriate factor loadings 
from default observations and their correlations. \citet{han&kang}
suggest introducing a further factor to model dependence of the economic factors without
violating the assumptions of the framework. Their approach is generalised in \citet{fischer2011modeling}.
Other authors \citep[e.g.][]{wang2015creditrisk+} propose extensions of the
CreditRisk$^+$ model that allow for realistic modelling of the dependencies between the economic
factors but renounce the analytic tractability of the original model. \citet{jakob2014quantifying}
discuss the impact on the loss distribution of choosing different factor dependence structures 
in extended versions of the CreditRisk$^+$ framework.

\section{Loss distributions conditional on defaults}\label{tas_sec_4}

The purpose of this section is to provide formulas for the portfolio
loss distribution conditional on defaults that can be represented in
similar terms as the unconditional loss distribution and hence be evaluated with the familiar
CreditRisk$^+$ algorithms. The following theorem -- a modification of
\citet[][Lemma~1]{Tasche2004} -- yields the foundation of the results.
Denote by $I(E)$ the indicator variable of the event $E$, i.e.\ $I(E; m) =1$
if $m\in E$ and $I(E;m) =0$ if $m\notin E$.
\begin{theorem}\label{tas_le_found}
Define the approximate default indicators $D_A$ as in Theorem \ref{tas_th_3}. Assume
that $A(1), \ldots, A(r)$ are obligors such that $A(i) \not= A(j)$ for $i \not= j$. Under
the assumptions and with the notation of Theorem \ref{tas_th_3} then we have
\begin{equation}\label{eq:fundamental}
  \mathbb{E}\Bigl[ I(X = x) \prod\limits_{i=1}^r D_{A(i)}\Bigr]\,=\,
\mathbb{E}\Bigl[ I\Bigl(X = x-\sum\limits_{j=1}^r \mathcal{E}_{A(j)}\Bigr) \prod\limits_{i=1}^r p_{A(i)}^S\Bigr]
\end{equation}
for any non-negative integer $x$, where the random variables
$\mathcal{E}_{A(1)}, \ldots, \mathcal{E}_{A(r)}$ on 
the right-hand side of \eqref{eq:fundamental} are independent of the loss variable $X$ and the default 
intensities $p^S_{A(i)}$.
\end{theorem}
\textbf{Proof.} We provide the proof only for the case $r = 2$ as the proof for general $r$ 
is not much different but the notation would be more cumbersome. 
Hence assume that two obligors $A(1) \not= A(2)$ have been selected. 
The assumptions on independence and conditional independence from Theorem~\ref{tas_th_3} then
imply
\begin{eqnarray*}
\lefteqn{\mathbb{E}\bigl[ D_{A(1)}\,D_{A(2)}\,I(X = x)\bigr]}\notag\\
& = &\sum_{k_1=1}^\infty \sum_{k_2=1}^\infty k_1\,k_2\,
\mathbb{P}\Big[D_{A(1)} = k_1, \,D_{A(2)} = k_2,
\sum_{\substack{B\not= A(1),\\ B\not=A(2)}} Y_B +
\sum\limits_{i=1}^{k_1}\mathcal{E}_{A(1),i} +
\sum\limits_{j=1}^{k_2}\mathcal{E}_{A(2),j} = x\Big]\\
& = &\sum_{k_1=1}^\infty \sum_{k_2=1}^\infty k_1\,k_2\,
\mathbb{E}\bigg[\frac{(p_{A(1)}^S)^{k_1}}{k_1!}\,e^{-p_{A(1)}^S}\,
\frac{(p_{A(2)}^S)^{k_2}}{k_2!}\,e^{-p_{A(2)}^S} \\
& \ & \quad\quad\quad\times \,\mathbb{P}\Big[\sum_{\substack{B\not= A(1),\\ B\not=A(2)}} Y_B +
\sum\limits_{i=1}^{k_1}\mathcal{E}_{A(1),i} +
\sum\limits_{j=1}^{k_2}\mathcal{E}_{A(2),j} = x\,|\,\mathbf{S}\Big]\bigg]\\
& = &\sum_{k_1=0}^\infty \sum_{k_2=0}^\infty 
\mathbb{E}\bigg[p_{A(1)}^S\,p_{A(2)}^S\\
& \ & \quad\quad\quad\times \,
\mathbb{P}\Big[D_{A(1)} = k_1, \,D_{A(2)} = k_2, \sum_{\substack{B\not= A(1),\\ B\not=A(2)}} Y_B +
\sum\limits_{i=1}^{k_1+1}\mathcal{E}_{A(1),i} +
\sum\limits_{j=1}^{k_2+1}\mathcal{E}_{A(2),j} = x\,|\,\mathbf{S}\Big]\bigg]\\
& = &
\mathbb{E}\left[ I\Bigl(X = x- \mathcal{E}_{A(1)}-\mathcal{E}_{A(2)}\Bigr)\, p_{A(1)}^S\,p_{A(2)}^S\right],
\end{eqnarray*}
as stated in \eqref{eq:fundamental}. \hfill q.e.d.

As the variable $D_A$ approximates obligor $A$'s default indicator the
conditional expectation $\mathbb{E}[D_A\,|\,X=x]$ can be interpreted as an approximation of the
conditional probability of obligor $A$'s default given that the portfolio loss
$X$ assumes the value $x$. \citet[][Corollary~1]{Tasche2004} observed the following
result for $\mathbb{E}[D_A\,|\,X=x]$. It can be readily derived from Theorem~\ref{tas_le_found}.

\textbf{Notation.} \emph{For any positive integers $i \le n$ define the \emph{$n$-dimensional $i$-th unit vector} 
$e_i^{(n)}$ by }
\begin{equation*}
	e_i^{(n)} = (\underbrace{0, \ldots, 0}_{i-1\text{\ times}}, 1, 
	\underbrace{0, \ldots, 0}_{n-i\text{\ times}}).
\end{equation*}
\emph{Where the dimension is known from the context we write $e_i = e_i^{(n)}$ for short.}

\begin{corollary}[Probability of default conditional on portfolio loss]\label{tas_co_1}\ \\
Adopt the setting and the notation of Theorem~\ref{tas_th_3} and Theorem~\ref{tas_le_found}. 
Write $\mathbb{P}_\alpha[X \in
\,\cdot\,]$ for $\mathbb{P}[X \in
\,\cdot\,]$ in order to express the dependence\footnote{%
Of course, the distribution also depends on $\mu_0$, $\mathcal{Q}_0, \ldots,
\mathcal{Q}_N$, and $\delta_1, \ldots, \delta_N$. However, these input
parameters are considered constant in Corollary \ref{tas_co_1}.}%
of the portfolio loss distribution upon the exponents $\alpha = (\alpha_1,
\ldots, \alpha_N)$ in (\ref{tas_eq_gen}). Assume that $x$ is an integer such that
$\mathbb{P}_{\alpha}[X = x] > 0$.
Then, in the
CreditRisk$^+$ framework, the conditional probability of obligor $A$'s default
given that the portfolio loss $X$ assumes the value $x$ can be approximated by
\begin{equation}
\mathbb{E}[D_A\,|\,X=x]\  =\ p_A\,
 \frac{w_{A 0}\,\mathbb{P}_{\alpha}[X = x-\widetilde{\mathcal{E}}_A] +
\sum_{j=1}^N w_{A j}\, \mathbb{P}_{\alpha + e_j}[X = x -
\widetilde{\mathcal{E}}_A]}{\mathbb{P}_{\alpha}[X = x]}, \label{tas_cond_prob}
\end{equation}
where $\widetilde{\mathcal{E}}_A$ stands for a random variable that has the
same distribution as $\mathcal{E}_A$ but is independent of $X$.
\end{corollary}
Intuitively, one might think that $\mathbb{P}[D_A > 0\,|\,X=x]$ would be a better approximation of
the conditional probability of default of obligor $A$ than $\mathbb{E}[D_A\,|\,X=x]$. However, there is
no such relatively simple representation of $\mathbb{P}[D_A > 0\,|\,X=x]$ as \eqref{tas_cond_prob} is 
for $\mathbb{E}[D_A\,|\,X=x]$. 
Moreover, by the assumption on the conditional Poisson distribution of $D_A$ we have
\begin{equation}
	\mathbb{E}\bigl[\mathbb{P}[D_A > 0\,|\,X]\bigr] =  \mathbb{P}[D_A > 0] < 
	p_A = \mathbb{E}\bigl[\mathbb{E}[D_A\,|\,X]\bigr].
\end{equation}
Hence the bias of $\mathbb{P}[D_A > 0\,|\,X=x]$ with respect to $\mathbb{P}[A\
\text{defaults}\,|\,X=x]$ is likely to be greater than the bias of $\mathbb{E}[D_A\,|\,X=x]$.

The probabilities in the numerator of the right-hand side of (\ref{tas_cond_prob}) must be
calculated by convolution if the loss severities $\mathcal{E}_A$ are non-deterministic. 
In any case, Corollary \ref{tas_co_1} can be used for constructing
the portfolio loss distribution conditional on the default of an obligor.
Observe that by the very definition of conditional probabilities it follows that
\begin{equation}\label{tas_eq_default}
  \mathbb{P}[X = x\,|\,A\ \text{defaults}] \  =\ \mathbb{P}[A\ \text{defaults}\,|\,X=x]
  	\frac{\mathbb{P}[X = x]}{p_A}.
\end{equation}
Since by Corollary \ref{tas_co_1} an approximation for $\mathbb{P}[A\
\text{defaults}\,|\,X=x]$ is provided, the term-wise comparison of
(\ref{tas_cond_prob}) and (\ref{tas_eq_default}) yields
\begin{equation}\label{tas_eq_cond_distr}
\mathbb{P}_\alpha[X = x\,|\,A\
  \text{defaults}] \ \approx \
w_{A 0}\,\mathbb{P}_{\alpha}[X = x-\widetilde{\mathcal{E}}_A]\, + \sum_{j=1}^N
\;w_{A j}\, \mathbb{P}_{\alpha + e_j}[X = x - \widetilde{\mathcal{E}}_A].
\end{equation}
Note that according to (\ref{tas_eq_cond_distr}), the conditional distribution
$\mathbb{P}_\alpha[X = \,\cdot\,|\,A\ \text{defaults}]$ of the portfolio loss
$X$ given that $A$ defaults may be computed as a weighted mean of stressed 
portfolio loss distributions. The stresses are expressed by the exponents $\alpha_j + 1$ in the
generating functions of $\mathbb{P}_{\alpha + e_j}[X = \,\cdot\,]$, $j=1, \ldots, N$. 
In actuarial terms, incrementing the size parameter of a negative binomial 
claim number distribution (cf.\ Remark~\ref{tas_rm_0}) means to give the claim number distribution a heavier tail. Hence,
this way the number of claims (sector-related defaults in CreditRisk$^+$ terms) tends to
be larger after the stress was applied. No change due to stress, however, occurs to the
sector loss severity distributions as characterised by the sector polynomials $\mathcal{Q}_j$.
This is no surprise as the loss severities in the setting of this paper are assumed to be
independent of the economic factors that drive the sector default frequencies.

\pagebreak
\begin{samepage}
\begin{remark}\label{rm:deduction}\ 
\begin{enumerate}
	\item[(i)] By (\ref{tas_eq_cond_distr}) stressed portfolio loss distributions can be
evaluated, conditional on the scenarios that single obligors have defaulted.
If, for instance, the portfolio Value-at-Risk changes dramatically when
obligor $A$'s default is assumed, then one may find that the portfolio
depends too strongly upon $A$'s condition.
	\item[(ii)] Equation (\ref{tas_eq_cond_distr}) reflects a write-off or special provision due 
to obligor $A$'s default. This is a consequence of the fact that on the right-hand side of
the equation loss distributions of the shape $X + \widetilde{\mathcal{E}}_A$ appear, thus implying
that losses $X$ are added to a loss socket $\widetilde{\mathcal{E}}_A$ caused by obligor $A$'s
first default. However, usually in banks occurred losses are not taken into account for 
the determination of risk metrics (like quantiles as defined by \eqref{tas_eq_3}) but are deducted
from the banks available capital buffer. In that sense (\ref{tas_eq_cond_distr}) does not 
appropriately reflect banks' practice.
	\item[(iii)] To deal with the issue observed in (ii), note that Theorem~\ref{tas_th_3} and
	Corollary~\ref{tas_co_1} also can be applied to the case $\mathcal{E}_A = 0$. In particular,
	dependencies within the portfolio are then still adequately reflected by obligor
	$A$'s conditional default intensity $p_A^S$. While $\mathcal{E}_A = 0$ in Theorem~\ref{tas_th_3} effectively
	eliminates any impact of obligor $A$ on the unconditional portfolio loss distribution, \eqref{tas_eq_cond_distr} 
	clearly
	demonstrates the impact of the dependence between $A$ and the rest of the portfolio on the conditional
	portfolio loss distribution. 
\end{enumerate}
\end{remark}
\end{samepage}
While Theorem~\ref{tas_le_found} can be used to study the portfolio loss distributions
conditional on any number of defaults, we confine ourselves in the following 
corollary and its consequences to considering only
the case of two defaults as we already did in the proof of Theorem~\ref{tas_le_found}.
The formulas for conditioning on three or more defaults can be derived in the same way 
as the formula for the case of two defaults. The cases of three or more defaults, however,
are notationally and computationally much more inconvenient, presumably much less relevant for practice, and 
do not add much more theoretical insight compared to the case of two defaults.
\begin{corollary}[Joint probability of default conditional on portfolio loss]\label{co:joint}\ \\
Adopt the setting and the notation of Corollary~\ref{tas_co_1}. Let $A(1) \not= A(2)$ denote
two obligors who have been selected in advance. Assume that $x$ is an integer such that
$\mathbb{P}_{\alpha}[X = x] > 0$.
Then, in the
CreditRisk$^+$ framework, the conditional joint probability of obligor $A(1)$'s 
and obligor $A(2)$'s default
given that the portfolio loss $X$ assumes the value $x$ may be approximated by
\begin{multline}
\mathbb{E}[D_{A(1)}\,D_{A(2)}\,|\,X=x]\ =\ 
\frac{p_{A(1)}\,p_{A(2)}}{\mathbb{P}_{\alpha}[X = x]}
\Big(w_{A(1) 0}\,w_{A(2) 0}\,
	\mathbb{P}_{\alpha}[X = x-\widetilde{\mathcal{E}}_{A(1)}-\widetilde{\mathcal{E}}_{A(2)}]\ +\\
     \sum_{j=1}^N \Bigl(w_{A(1) 0}\,w_{A(2) j} + w_{A(1) j}\,w_{A(2) 0}\Bigr)\, 
		\mathbb{P}_{\alpha + e_j}[X = x - \widetilde{\mathcal{E}}_{A(1)}-\widetilde{\mathcal{E}}_{A(2)}]\ +\\
	 \sum_{j=1}^N w_{A(1) j}\,w_{A(2) j} \, \frac{\alpha_j+1}{\alpha_j}\,
		\mathbb{P}_{\alpha + 2 e_j}[X = x - \widetilde{\mathcal{E}}_{A(1)}-\widetilde{\mathcal{E}}_{A(2)}]\ +\\
 \sum_{i=1}^N \sum_{\substack{j=1, \\ j\not=i}}^N w_{A(1) i}\,w_{A(2) j} \, 
		\mathbb{P}_{\alpha + e_i + e_j}[X = x - \widetilde{\mathcal{E}}_{A(1)}-\widetilde{\mathcal{E}}_{A(2)}]			
		\Big)\label{tas_eq_CRplus_cont}
\end{multline}
where $\widetilde{\mathcal{E}}_A$ for $A=A(1)$ and $A=A(2)$ stands for a random variable that has the
same distribution as $\mathcal{E}_A$ but is independent of $X$.
\end{corollary}
While \eqref{tas_eq_CRplus_cont} in general looks like a straight-forward extension of \eqref{tas_cond_prob},
there is a subtle difference in the terms involving $\mathbb{P}_{\alpha + 2 e_j}[X = x - \widetilde{\mathcal{E}}_{A(1)}-\widetilde{\mathcal{E}}_{A(2)}]$ which reflect double stress in the same
sector. This double stress is enforced by the additional factors $\frac{\alpha_j+1}{\alpha_j} > 1$.

\textbf{Proof of Corollary \ref{co:joint}.} We derive
(\ref{tas_eq_CRplus_cont}) by comparing the coefficients of two power series.
The first one is $\mathbb{E}[D_{A(1)}\,D_{A(2)}\,z^X] = \sum_{k=0}^\infty
\mathbb{E}[D_{A(1)}\,D_{A(2)}\,I(X=k)]\,z^k$, the second one is an expression that is
equivalent to $\mathbb{E}[D_{A(1)}\,D_{A(2)}\,z^X]$ but involves generating functions similar
to (\ref{tas_eq_gen}).

Recall that we denote the generating function of $\mathcal{E}_A$ by $H_A(z)$.
By means of Theorem~\ref{tas_le_found} and the independence of the random 
exposures, we can compute
\begin{subequations}
\begin{eqnarray}
\mathbb{E}[D_{A(1)}\,D_{A(2)}\,z^X]
&=& \sum_{k=0}^\infty 
\mathbb{E}[p_{A(1)}^S\,p_{A(2)}^S\,I(X+\mathcal{E}_{A(1)}+\mathcal{E}_{A(2)}=k)]\,z^k\notag\\[1ex]
&=& \mathbb{E}[p_{A(1)}^S\,p_{A(2)}^S\,z^{X+\mathcal{E}_{A(1)}+\mathcal{E}_{A(2)}}]\notag\\[1ex]
&=& \mathbb{E}[p_{A(1)}^S\,p_{A(2)}^S\,z^X]\,
		\mathbb{E}[z^{\mathcal{E}_{A(1)}}]\,\mathbb{E}[z^{\mathcal{E}_{A(2)}}]\notag\\[1ex]
&=& \mathbb{E}[p_{A(1)}^S\,p_{A(2)}^S\,z^X]\,H_{A(1)}(z) \,H_{A(2)}(z).\label{eq:joint_1}
 \label{tas_eq_gen_Y}
\end{eqnarray}
Recall the definitions of the intensities $p_A^S$, the 
sector default intensities $\mu_k$ and the sector polynomials
$\mathcal{Q}_k$ from Theorem~\ref{tas_th_3}. 
By making use of the fact that the economic factors $(S_1, \ldots, S_N)$ are
Gamma-distributed with parameters $(\alpha_k, 1/\alpha_k)$, $k=1, \ldots, N$, and that $S_0=1$
we obtain for $\mathbb{E}[p_{A(1)}^S\,p_{A(2)}^S\,z^X]$ (cf.\ the proof of (3.25c) in
\citet{Tasche2004})
\begin{eqnarray}
\mathbb{E}[p_{A(1)}^S\,p_{A(2)}^S\,z^X]&=& \mathbb{E}\bigl[p_{A(1)}^S\,p_{A(2)}^S\,\mathbb{E}[z^X\,|\,S]\bigr]
\label{eq:joint_2}\\
&=& p_{A(1)}\,p_{A(2)} \sum_{i=0}^N \sum_{j=0}^N w_{A(1) i}\,w_{A(2) j}\,\mathbb{E}\bigl[ S_i\,S_j\,\prod_{k=0}^N
\exp\bigl(S_k\,\mu_k\,(\mathcal{Q}_k(z) - 1)\bigr)\bigr].\notag
\end{eqnarray}
\end{subequations}
Denote by
\begin{equation}
G^{(\alpha)}_X(z) \ =\ \sum_{k=0}^\infty \mathbb{P}_\alpha[X = k]\,z^k
\end{equation}
the generating function of $X$ according to (\ref{tas_eq_gen}) as a function of
the exponents $\alpha = (\alpha_1, \ldots, \alpha_N)$ on the right-hand side of
the equation as has
been explained in Corollary~\ref{tas_co_1}.
Observe then that
\begin{equation}\label{eq:gen_func}
\begin{split}
\mathbb{E}\bigl[ S_0^2\prod_{k=0}^N
\exp\bigl(S_k\,\mu_k\,(\mathcal{Q}_k(z) - 1)\bigr)\bigr] & = G^{(\alpha)}_X(z)\\
\mathbb{E}\bigl[ S_0\,S_j\,\prod_{k=0}^N
\exp\bigl(S_k\,\mu_k\,(\mathcal{Q}_k(z) - 1)\bigr)\bigr]	& = G^{(\alpha+e_j)}_X(z), \ j \ge 1\\
\mathbb{E}\bigl[ S_i\,S_j\,\prod_{k=0}^N
\exp\bigl(S_k\,\mu_k\,(\mathcal{Q}_k(z) - 1)\bigr)\bigr] & = G^{(\alpha+e_i+e_j)}_X(z), \ i\not= j\\
\mathbb{E}\bigl[ S_j^2\,\prod_{k=0}^N
\exp\bigl(S_k\,\mu_k\,(\mathcal{Q}_k(z) - 1)\bigr)\bigr] & = 
\frac{\alpha_j+1}{\alpha_j}\,G^{(\alpha+2\,e_j)}_X(z), \ j \ge 1.
\end{split}
\end{equation}
Note that $G^{(\alpha)}_X(z)\,\,H_{A(1)}(z) \,H_{A(2)}(z)$ 
is the generating function of the sequence $\mathbb{P}_\alpha[X+\widetilde{\mathcal{E}}_{A(1)}+
\widetilde{\mathcal{E}}_{A(2)} =
0], \mathbb{P}_\alpha[X+\widetilde{\mathcal{E}}_{A(1)}+\widetilde{\mathcal{E}}_{A(2)} = 1], \ldots$ (i.e.\ of the distribution
of $X+\widetilde{\mathcal{E}}_{A(1)}+\widetilde{\mathcal{E}}_{A(2)}$).
Combining this observation with (\ref{eq:joint_1}), (\ref{eq:joint_2}), and  \eqref{eq:gen_func} implies 
(\ref{tas_eq_CRplus_cont}) by power series comparison. \hfill q.e.d.

As Corollary \ref{tas_co_1} can be used for constructing
the portfolio loss distribution conditional on the default of one obligor,
Corollary~\ref{co:joint} can be used for the portfolio loss distribution
conditional on the joint default of two obligors.
Again by the definition of conditional probabilities it follows that
\begin{multline}\label{eq:joint_defs}
   \mathbb{P}[X = x\,|\,\text{$A(1)$ and $A(2)$ default}] \  =\ \\
   \mathbb{P}[\text{$A(1)$ and $A(2)$ default}\,|\,X=x]
  	\frac{\mathbb{P}[X = x]}{\mathbb{P}[\text{$A(1)$ and $A(2)$ default}]}.
\end{multline}
Since by Corollary \ref{co:joint} an approximation for 
$\mathbb{P}[\text{$A(1)$ and $A(2)$ default}\,|\,X=x]$ is provided, the term-wise comparison of
(\ref{tas_eq_CRplus_cont}) and (\ref{eq:joint_defs}) yields
\begin{subequations}
\begin{multline}
\mathbb{P}_\alpha[X = x\,|\,\text{$A(1)$ and $A(2)$ default}]
 \ \approx \\
	\frac{p_{A(1)}\,p_{A(2)}}{\mathbb{P}[\text{$A(1)$ and $A(2)$ default}]}\, \Big(w_{A(1) 0}\,w_{A(2) 0}\,
	\mathbb{P}_{\alpha}[X = x-\widetilde{\mathcal{E}}_{A(1)}-\widetilde{\mathcal{E}}_{A(2)}]\ +\\ 
	 \sum_{j=1}^N (w_{A(1) 0}\,w_{A(2) j} + w_{A(1) j}\,w_{A(2) 0})\, 
		\mathbb{P}_{\alpha + e_j}[X = x - \widetilde{\mathcal{E}}_{A(1)}-\widetilde{\mathcal{E}}_{A(2)}]\ +\\
	 \sum_{j=1}^N w_{A(1) j}\,w_{A(2) j} \, \frac{\alpha_j+1}{\alpha_j}\,
		\mathbb{P}_{\alpha + 2 e_j}[X = x - \widetilde{\mathcal{E}}_{A(1)}-\widetilde{\mathcal{E}}_{A(2)}]\ +\\
	 \sum_{i=1}^N \sum_{\substack{j=1,\\ j\not=i}}^N w_{A(1) i}\,w_{A(2) j} \, 
		\mathbb{P}_{\alpha + e_i + e_j}[X = x - \widetilde{\mathcal{E}}_{A(1)}-\widetilde{\mathcal{E}}_{A(2)}]			
		\Big) \label{eq:cond_joint}
\end{multline}
Making use of the well-known result \citep[see][Section 2.3]{Gundlach2004}
\begin{equation}\label{eq:Gundlach}
	\mathbb{E}[D_{A(1)}\,D_{A(2)}] \ = \ p_{A(1)}\,p_{A(2)} \left(1 + 
	\sum_{k=1}^N \frac{w_{A(1) k}\,w_{A(2) k}}{\alpha_k}\right),\ A(1) \not= A(2), 
\end{equation}
\eqref{eq:cond_joint} can be slightly simplified to
\begin{eqnarray}
\lefteqn{\mathbb{P}_\alpha[X = x\,|\,\text{$A(1)$ and $A(2)$ default}]}\notag\\[1ex]
 & \approx &
	\frac{1}{1 + 
	\sum_{k=1}^N \frac{w_{A(1) k}\,w_{A(2) k}}{\alpha_k}}\, \Big(w_{A(1) 0}\,w_{A(2) 0}\,
	\mathbb{P}_{\alpha}[X = x-\widetilde{\mathcal{E}}_{A(1)}-\widetilde{\mathcal{E}}_{A(2)}]\ +\notag\\ 
	& \quad &  \sum_{j=1}^N \Bigl(w_{A(1) 0}\,w_{A(2) j} + w_{A(1) j}\,w_{A(2) 0}\Bigr)\, 
		\mathbb{P}_{\alpha + e_j}[X = x - 
			\widetilde{\mathcal{E}}_{A(1)}-\widetilde{\mathcal{E}}_{A(2)}]\ + \notag\\
	& \quad &  \sum_{j=1}^N w_{A(1) j}\,w_{A(2) j} \, \frac{\alpha_j+1}{\alpha_j}\,
		\mathbb{P}_{\alpha + 2 e_j}[X = x - 
			\widetilde{\mathcal{E}}_{A(1)}-\widetilde{\mathcal{E}}_{A(2)}]\ + \notag\\
	& \quad &  \sum_{i=1}^N \sum_{\substack{j=1,\\ j\not=i}}^N w_{A(1) i}\,w_{A(2) j} \, 
		\mathbb{P}_{\alpha + e_i + e_j}
		[X = x - \widetilde{\mathcal{E}}_{A(1)}-\widetilde{\mathcal{E}}_{A(2)}]	\Big). \label{eq:cond_joint_simp}
\end{eqnarray}
\end{subequations}
Comments similar to the comments on (\ref{tas_eq_cond_distr}) also apply to \eqref{eq:cond_joint_simp}. 
The conditional distribution 
$\mathbb{P}_\alpha[X = x\,|\,\text{$A(1)$ and $A(2)$ default}]$ of the portfolio loss
$X$ given that obligors $A(1)$ and $A(2)$ default can be computed as a weighted mean of stressed or
double-stressed
portfolio loss distributions. The stresses, however, are not only expressed by the exponents $\alpha_j + 1$ 
and $\alpha_j + 2$ in the
generating functions of $\mathbb{P}_{\alpha + e_j}[X = \,\cdot\,]$ and 
$\mathbb{P}_{\alpha + e_i + e_j}[X = \,\cdot\,]$, $i, j=1, \ldots, N$, but also by the factors 
$\frac{\alpha_j+1}{\alpha_j} > 1$ appearing on the right-hand side of \eqref{eq:cond_joint_simp}. Obviously,
as a consequence of the $(N+1)^2$ terms on the right-hand side of \eqref{eq:cond_joint_simp} instead of
the only $N+1$ terms of the right-hand side of (\ref{tas_eq_cond_distr}), it is much more expensive to
calculate the loss distributions conditional on two defaults than to calculate the loss 
distributions conditional on simple defaults.

Observe that Remark~\ref{rm:deduction} also applies to \eqref{eq:cond_joint_simp}. Hence it makes
sense to do the calculations for \eqref{eq:cond_joint_simp} with loss severities $\mathcal{E}_{A(1)} = 0$
and $\mathcal{E}_{A(2)} = 0$ to reflect the risk management attitude not to take account of occurred losses
for the determination of living portfolio risk metrics.


\section{The `stressed probabilities of default' approach}
\label{sec:StressedParameters}

Under the CreditRisk$^+$ framework, equation \eqref{eq:cond_joint_simp} provides the algorithm needed 
for the calculation of the portfolio
loss distribution conditional on the default of two obligors. However, if $N$ denotes the number of economic
factors in the model, formula \eqref{eq:cond_joint_simp} requires the computation of $\frac{(N+1)\,(N+2)}{2}$
slightly different loss distributions which could be tedious if $N$ is large. In this section, therefore,
we look at the `cheaper' alternative approach where the loss distribution is calculated only once according
to Theorem~\ref{tas_th_3} and all parameters but the unconditional probabilities of default $p_A$ remain unchanged.
In this `stressed probabilities of default' approach the $p_A$ are replaced by probabilities of default conditional
on the default of the two obligors. The approach is based on the following three-events version of \eqref{eq:Gundlach}.

\begin{proposition}\label{tas_le_3defaults}
Define the approximate default indicators $D_A$ as in Theorem \ref{tas_th_3}. Assume
that $A(1), A(2), A(3)$ are obligors such that $A(i) \not= A(j)$ for $i \not= j$. Under
the assumptions and with the notation of Theorem \ref{tas_th_3} then we have
\begin{multline*}
  \mathbb{E}[D_{A(1)}\,D_{A(2)}\,D_{A(3)}]\ =\\ 
  p_{A(1)}\,p_{A(2)}\,p_{A(3)}\,
\Big(1 + 2 \sum_{k=1}^N \tfrac{w_{A(1) k}\,w_{A(2) k}\,w_{A(3) k}}{\alpha_k^2} 
+ \sum_{k=1}^N  \tfrac{w_{A(1) k}\,w_{A(2) k} + w_{A(1) k}\,w_{A(3) k} + w_{A(2) k}\,w_{A(3) k}}{\alpha_k} \Big).     
\end{multline*}
\end{proposition}

\textbf{Proof.}  The assumption on
the Poisson distribution of the $D_A$ conditional on the vector of economic factors $\mathbf{S} = 
(S_0, S_1, \ldots, S_N)$ implies $\mathbb{E}[D_A\,|\,\mathbf{S}] = p_A^S$ with $p_A^S$ defined as in 
Theorem~\ref{tas_th_3}. By the conditional independence of $A(1), A(2), A(3)$ therefore it follows that
\begin{align*}
\mathbb{E}[D_{A(1)}\,D_{A(2)}\,D_{A(3)}] & = \mathbb{E}[p_{A(1)}^S\,p_{A(2)}^S\,p_{A(3)}^S]\\
& =  p_{A(1)}\,p_{A(2)}\,p_{A(3)} \sum_{j=0}^N \sum_{k=0}^N \sum_{\ell=0}^N 
    w_{A(1) j}\,w_{A(2) k}\,w_{A(3) \ell}\, \mathbb{E}[S_j\,S_k\,S_\ell].
\end{align*}
Recall that by assumption we have $\mathbb{E}[S_k] = 1$ for all $k=1, \ldots, N$ and $\sum_{k=0}^N w_{A k} = 1$ for
all obligors $A$. This implies
\begin{align*}
\frac{\mathbb{E}[D_{A(1)}\,D_{A(2)}\,D_{A(3)}]}{p_{A(1)}\,p_{A(2)}\,p_{A(3)}} 
& =  1 +  \sum_{k=0}^N \sum_{\substack{\ell=0,\\ \ell\not=k}}^N 
    w_{A(1) k}\,w_{A(2) k}\,w_{A(3) \ell}\,\mathrm{var}[S_k] \\
& \qquad + \sum_{k=0}^N \sum_{\substack{\ell=0,\\ \ell\not=k}}^N 
    w_{A(1) k}\,w_{A(2) \ell}\,w_{A(3) k}\,\mathrm{var}[S_k] \\
& \qquad + \sum_{k=0}^N \sum_{\substack{\ell=0,\\ \ell\not=k}}^N 
    w_{A(1) \ell}\,w_{A(2) k}\,w_{A(3) k}\,\mathrm{var}[S_k] \\
& \qquad + \sum_{k=0}^N 
    w_{A(1) k}\,w_{A(2) k}\,w_{A(3) k}\,\bigl(\mathbb{E}[S_k^3]-1\bigr) \\
& =  1 +  \sum_{k=0}^N 
    w_{A(1) k}\,w_{A(2) k}\,(1-w_{A(3) k})\,\mathrm{var}[S_k] \\
& \qquad + \sum_{k=0}^N  
    w_{A(1) k}\,(1-w_{A(2) k})\,w_{A(3) k}\,\mathrm{var}[S_k] \\
& \qquad + \sum_{k=0}^N 
    (1-w_{A(1) k})\,w_{A(2) k}\,w_{A(3) k}\,\mathrm{var}[S_k] \\
& \qquad + \sum_{k=0}^N w_{A(1) k}\,w_{A(2) k}\,w_{A(3) k}\,\bigl(\mathbb{E}[S_k^3]-1\bigr).
\end{align*}
From the assumption that $S_k$ is Gamma-distributed with parameter vector $(\alpha_k, 1/\alpha_k)$, it follows
that $\mathrm{var}[S_k] = 1/\alpha_k$ and $\mathbb{E}[S_k^3]-1 = \frac{3\,\alpha_k+2}{\alpha_k^2}$. This
implies the assertion. \hfill q.e.d.

Since in the CreditRisk$^+$ framework the default indicator for an obligor $A$ is approximated by the
conditional Poisson variable $D_A$, the joint probability of default 
$\mathbb{P}[A(1)\,\text{and}\,A(2)\,\text{and}\,A(3)\,\text{default}]$ of three obligors is
approximated by
$$\mathbb{P}[A(1)\,\text{and}\,A(2)\,\text{and}\,A(3)\,\text{default}]
\ \approx\ \mathbb{E}[D_{A(1)}\,D_{A(2)}\,D_{A(3)}].$$
Hence Proposition~\ref{tas_le_3defaults} and \eqref{eq:Gundlach} provide us with a simple approximation
formula for one obligor's probability of default conditional on two other obligors' joint default:
\begin{multline}\label{eq:ConditionalOn2}
    \mathbb{P}[B\,\text{defaults}\,|\,A(1)\,\text{and}\,A(2)\,\text{default}] \ \approx \\ 
    \frac{p_B}{1 + 
	\sum_{k=1}^N \frac{w_{A(1) k}\,w_{A(2) k}}{\alpha_k}}
     \Big(1 + 2 \sum_{k=1}^N \tfrac{w_{B k}\,w_{A(1) k}\,w_{A(2) k}}{\alpha_k^2} 
+ \sum_{k=1}^N  \tfrac{w_{B k}\,w_{A(1) k} + w_{B k}\,w_{A(2) k} + w_{A(1) k}\,w_{A(2) k}}{\alpha_k}\Big),
\end{multline} 
for any three different obligors $B$, $A(1)$ and $A(2)$. Thanks to Proposition~\ref{tas_le_3defaults} and
Equation~\eqref{eq:ConditionalOn2}, 
we can describe in precise technical terms the two above mentioned approaches 
to the calculation
of the loss distribution conditional on two defaults.  

\begin{definition}\label{de:measures}
In the setting and with the notation of Theorem~\ref{tas_th_3}, assume that there are two obligors $A(1)$ and $A(2)$ 
with exposures $\mathcal{E}_{A(1)} = \mathcal{E}_{A(2)} = 0$. Call this setting the \emph{two defaults scenario}.
\begin{itemize}
\item[(i)] The portfolio loss distribution defined by the right-hand-side of \eqref{eq:cond_joint_simp} is called
the \emph{`two defaults scenario' loss distribution}.
\item[(ii)] Replace in (i) of Theorem~\ref{tas_th_3} the probabilities of default $p_A$ by the 
conditional probabilities of default 
$\mathbb{P}[A\,\mathrm{defaults}\,|\,A(1)\,\mathrm{and}\,A(2)\,\mathrm{default}]$ 
as given by \eqref{eq:ConditionalOn2} and
keep all other parameters in the theorem unchanged. The resulting portfolio loss distribution is called 
\emph{`stressed probabilities of default' loss distribution}.
\end{itemize}
\end{definition}
Intuitively, it is clear that the expected value $\mathbb{E}[X]$ of the 
portfolio loss $X$ should be the same under
both loss distributions from Definition~\ref{de:measures}. However, since the right-hand-sides of 
both \eqref{eq:cond_joint_simp} and \eqref{eq:ConditionalOn2}
are only approximations to the conditional probabilities on the left-hand-sides of the equations, 
the fact that the two
expected values are equal must be formally proven.

\begin{proposition}\label{pr:EqualMeans}
Under the `two defaults scenario', denote by $\mathbb{P}_{\mathrm{Two}}$ 
the distribution of Definition~\ref{de:measures}~(i) and by 
$\mathbb{P}_{\mathrm{Prob}}$ the distribution of Definition~\ref{de:measures}~(ii). Then it holds that
$\mathbb{E}_{\mathrm{Two}}[X] = \mathbb{E}_{\mathrm{Prob}}[X]$.
\end{proposition}
\textbf{Proof.} For the sake of a clear notation, we denote all obligors but 
$A(1)$ and $A(2)$ with the letter $B$. 
Under the independence assumptions of Theorem~\ref{tas_th_3}, by construction of $\mathbb{P}_{\mathrm{Prob}}$
Equation~\eqref{tas_eq_loss} implies that
\begin{multline}\label{eq:Prob}
\mathbb{E}_{\mathrm{Prob}}[X]  =  \sum_B \mathbb{E}_{\mathrm{Prob}}[D_B]\,\mathbb{E}[\mathcal{E}_B]
= \sum_B \mathbb{P}[B\,\text{defaults}\,|\,A(1)\,\text{and}\,A(2)\,\text{default}]\,\mathbb{E}[\mathcal{E}_B] \\
 = \frac{\sum_B p_B\,\mathbb{E}[\mathcal{E}_B]\,\left(1 + 2 \sum_{k=1}^N \tfrac{w_{B k}\,w_{A(1) k}\,w_{A(2) k}}{\alpha_k^2} 
+ \sum_{k=1}^N  \tfrac{w_{B k}\,w_{A(1) k} + w_{B k}\,w_{A(2) k} + w_{A(1) k}\,w_{A(2) k}}{\alpha_k}\right)}{1 + \sum_{k=1}^N \frac{w_{A(1) k}\,w_{A(2) k}}{\alpha_k}}.
\end{multline}
For $\mathbb{E}_{\mathrm{Two}}[X]$, we obtain from \eqref{eq:cond_joint_simp} that
\begin{multline}\label{eq:Two}
\mathbb{E}_{\mathrm{Two}}[X]  = \left(1 + \sum_{k=1}^N \frac{w_{A(1) k}\,w_{A(2) k}}{\alpha_k}\right)^{-1}\,
\Big(w_{A(1) 0}\,w_{A(2) 0}\,
	\mathbb{E}_{\alpha}[X]\ + \\ \sum_{j=1}^N \bigl(w_{A(1) 0}\,w_{A(2) j} + w_{A(1) j}\,w_{A(2) 0}\bigr)\, 
		\mathbb{E}_{\alpha + e_j}[X]\ +\\
	 \sum_{j=1}^N w_{A(1) j}\,w_{A(2) j} \, \frac{\alpha_j+1}{\alpha_j}\,
		\mathbb{E}_{\alpha + 2 e_j}[X]
	+ \sum_{i=1}^N \sum_{\substack{j=1,\\ j\not=i}}^N w_{A(1) i}\,w_{A(2) j} \, 
		\mathbb{E}_{\alpha + e_i + e_j}[X]\Big).
\end{multline}
The distributions of $X$ referred to in the expected values on the right-hand-side of \eqref{eq:Two} are 
specified by the generating function \eqref{tas_eq_gen}. As explained in Remark~\ref{tas_rm_0} (ii), for instance
the distribution of $X$ under $\mathbb{P}_{\alpha + 2 e_j}$ is given by the convolution of a compound Poisson
distribution with expected value $\sum_B w_{B 0}\,p_B\,\mathbb{E}[\mathcal{E}_B]$ and $N$ compound negative 
binomial distributions with expected values 
\begin{align*}
\frac{\alpha_k\,\delta_k}{1-\delta_k}\,\frac{\sum_B w_{B k}\,p_B\,\mathbb{E}[\mathcal{E}_B]}{\mu_k} &\ =\ 
\sum_B w_{B k}\,p_B\,\mathbb{E}[\mathcal{E}_B], & \quad & k=1, \ldots, N, k \not= j,  \\
\frac{(\alpha_j+2)\,\delta_j}{1-\delta_j}\,\frac{\sum_B w_{B j}\,p_B\,\mathbb{E}[\mathcal{E}_B]}{\mu_j} &\ =\
\frac{\alpha_j+2}{\alpha_j}\,\sum_B w_{B j}\,p_B\,\mathbb{E}[\mathcal{E}_B], & \quad & k = j. 
\end{align*}
Substituting all these expected values into \eqref{eq:Two} and taking into account that $\sum_{k=0}^N w_{B k} = 1$
for all $B$ gives
\begin{eqnarray*}
\lefteqn{\left(1 + \sum_{k=1}^N \frac{w_{A(1) k}\,w_{A(2) k}}{\alpha_k}\right) \mathbb{E}_{\mathrm{Two}}[X]}\notag\\
& = & w_{A(1) 0}\,w_{A(2) 0}\,\sum_B p_B\,\mathbb{E}[\mathcal{E}_B]\ + \notag\\
&  & \sum_{j=1}^N \bigl(w_{A(1) 0}\,w_{A(2) j} + w_{A(1) j}\,w_{A(2) 0}\bigr) \left(\sum_B p_B\,\mathbb{E}[\mathcal{E}_B] 
    + \tfrac{1}{\alpha_j}\,\sum_B p_B\,w_{B j}\,\mathbb{E}[\mathcal{E}_B] \right)\ + \notag\\
& & \sum_{j=1}^N w_{A(1) j}\,w_{A(2) j} \, \tfrac{\alpha_j+1}{\alpha_j} \left(\sum_B p_B\,\mathbb{E}[\mathcal{E}_B] 
    + \tfrac{2}{\alpha_j}\,\sum_B p_B\,w_{B j}\,\mathbb{E}[\mathcal{E}_B] \right)\ + \notag\\
& & \sum_{i=1}^N \sum_{\substack{j=1,\\ j\not=i}}^N w_{A(1) i}\,w_{A(2) j}
    \left(\sum_B p_B\,\mathbb{E}[\mathcal{E}_B] 
    + \tfrac{1}{\alpha_j}\,\sum_B p_B\,w_{B j}\,\mathbb{E}[\mathcal{E}_B] +
    \tfrac{1}{\alpha_i}\,\sum_B p_B\,w_{B i}\,\mathbb{E}[\mathcal{E}_B]\right) \notag\\
& = & \sum_B p_B\,\mathbb{E}[\mathcal{E}_B] + \sum_{j=1}^N \bigl(w_{A(1) 0}\,w_{A(2) j} + w_{A(1) j}\,w_{A(2) 0}\bigr) \, 
    \tfrac{1}{\alpha_j}\,\sum_B p_B\,w_{B j}\,\mathbb{E}[\mathcal{E}_B]\ + \notag\\
& & \sum_{j=1}^N w_{A(1) j}\,w_{A(2) j} \, \left(\tfrac{1}{\alpha_j} \sum_B p_B\,\mathbb{E}[\mathcal{E}_B] 
    + \tfrac{2\,(\alpha_j+1)}{\alpha_j^2}\,\sum_B p_B\,w_{B j}\,\mathbb{E}[\mathcal{E}_B] \right)\ + \notag\\
& & \sum_{i=1}^N \sum_{\substack{j=1,\\ j\not=i}}^N w_{A(1) i}\,w_{A(2) j}
    \left(\tfrac{1}{\alpha_j}\,\sum_B p_B\,w_{B j}\,\mathbb{E}[\mathcal{E}_B] +
    \tfrac{1}{\alpha_i}\,\sum_B p_B\,w_{B i}\,\mathbb{E}[\mathcal{E}_B]\right). 
\end{eqnarray*}
Some algebra shows that the sum of the terms after the last "$=$" sign divided by the 
factor $1 + \sum_{k=1}^N \frac{w_{A(1) k}\,w_{A(2) k}}{\alpha_k}$ is equal to the right-hand-side of \eqref{eq:Prob}.
\hfill q.e.d.

Why are the `two defaults scenario' loss distribution $\mathbb{P}_{\mathrm{Two}}$ and the 
`stressed probabilities of default' loss distribution $\mathbb{P}_{\mathrm{Prob}}$ of Definition~\ref{de:measures}
 different despite the first order equality of the two
demonstrated in Proposition~\ref{pr:EqualMeans}? They differ because $\mathbb{P}_{\mathrm{Prob}}$
does not account for correct conditional joint probabilities of default for two or more obligors. Nonetheless, it is 
not clear how much the two loss distribution can differ, 
given that their first moments are equal. 
In the next section, we will consider two simple numerical examples to compare the two loss distributions
and assess how different they may be.

Another question refers to the nature of the input parameters $p_A$ in Theorem~\ref{tas_th_3}, i.e.\ the unconditional
PDs of the obligors in the portfolio. In principle, these PDs should be `through-the-cycle' (TTC) PDs\footnote{%
See \citet{jobst2012capital} for a formal definition of TTC PDs and a discussion of TTC v.\ PIT (point-in-time)
PDs.} 
in the CreditRisk$^+$ framework. Does is then make sense to use conditional PDs as input parameters to the model
as in the `stressed PDs' approach? Actually, this question misses the point. For `stressed PDs' only is meant to
be a technical workaround for more demanding approaches like Monte-Carlo simulation and
the calculation of the proper loss distribution conditional on two defaults (the `two defaults scenario'
distribution).


\section{Numerical examples}
\label{sec:Examples}

The first example we consider is a homogeneous portfolio with a one-factor dependence structure. For the factor, 
we choose a standard deviation of 0.8 which according to \citet{Merino&Nyfeler2004} is in the centre of the range
of observable default rate volatilities. Since the factor is assumed to be Gamma-distributed with mean~1, a standard
deviation of 0.8 implies that the factor is Gamma-distributed with parameters $(\alpha, \beta) =
(1/0.8^2, 0.8^2) = (1.5625, 0.64)$.
\begin{example}\label{ex:hom}
We assume the setting of Theorem~\ref{tas_th_3} with the following specifics:
\begin{itemize}
\item There are $n$ obligors $B_1, \ldots, B_n$ all with PD~$p = 1\%$. There is one economic factor $S$ such that
the conditional Poisson distribution of the default indicator $D_{B_i}$ is given by the intensity
$p^S_{B_i} = p\,S$ for all $i = 1, \ldots, n$.
\item Two further obligors $A_1$ and $A_2$ with the same characteristics as the other obligors are known to have 
defaulted.
\item The factor $S$ is Gamma-distributed with parameters $(1.5625, 0.64) = (\alpha, 1/\alpha)$.
\item The exposure to each of the obligors but the defaulters is 1. Hence we have $H_{B_i}(z) = z$ for the
generating functions of the exposures for all $i$. The exposures to the two defaulters $A_1$ and $A_2$ are 0.
\end{itemize}
Having all exposures equal to 1 means that in this case the portfolio `loss' distribution is actually the 
distribution of the number of defaults in the portfolio, i.e.\ we have
$$X \ = \ \sum_{i=1}^n D_{B_i}.$$
By Remark~\ref{tas_rm_0} (ii), it follows that in the CreditRisk$^+$ framework 
the unconditional distribution of $X$ is 
negative binomial and as such given by 
$$\mathbb{P}[X = x] \ = \ \frac{\Gamma(\alpha+x)}{\Gamma(\alpha)\,x!} \left(1-\frac{n\,p}{n\,p+\alpha}\right)^\alpha
    \left(\frac{n\,p}{n\,p+\alpha}\right)^x, \quad x = 0, 1, 2, \ldots.$$
Equation~\eqref{eq:cond_joint_simp} implies that the distribution of $X$ conditional on the default of 
$A_1$ and $A_2$ is approximated by
\begin{multline*}
$$\mathbb{P}[X = x\,|\, A_1\,\mathrm{and}\,A_2\,\mathrm{default}] \ =  
\\ \frac{\Gamma(\alpha+2+x)}{\Gamma(\alpha+2)\,x!} \left(1-\frac{n\,p}{n\,p+\alpha}\right)^{\alpha+2}
    \left(\frac{n\,p}{n\,p+\alpha}\right)^x, \quad x = 0, 1, 2, \ldots.
\end{multline*}
Again by Remark~\ref{tas_rm_0} (ii), it follows that the `stressed probabilities of default' distribution $\mathbb{Q}$
of $X$ in the sense of Definition~\ref{de:measures} is given by $q = \frac{p\,(\alpha+2)}{\alpha}$ and
$$\mathbb{Q}[X = x] \ = \ \frac{\Gamma(\alpha+x)}{\Gamma(\alpha)\,x!} \left(1-\frac{n\,q}{n\,q+\alpha}\right)^\alpha
    \left(\frac{n\,q}{n\,q+\alpha}\right)^x, \quad x = 0, 1, 2, \ldots.$$
\end{example}
\begin{figure}[t!p]
\caption{\small{}Unconditional, conditional on two defaults and 
`stressed input PDs' distributions of the number of defaults in portfolio of 100 obligors. All obligors have
unconditional PD~1\%. The model is one-factor CreditRisk$^+$ with factor standard deviation 0.8.}
\label{fig:tasche}
\begin{center}
\ifpdf
	\includegraphics[width=15cm]{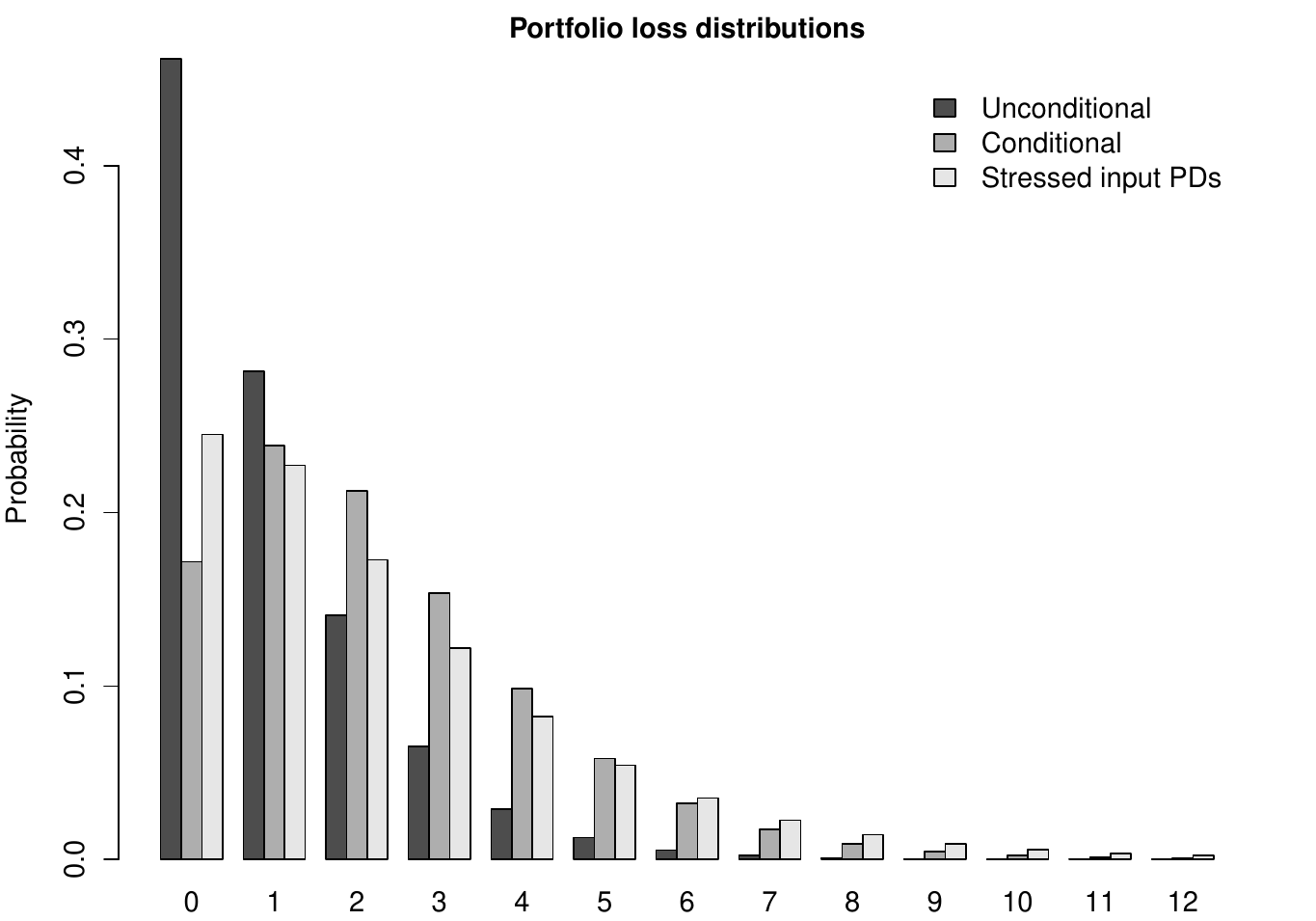}
\fi
\end{center}
\end{figure}
Figure~\ref{fig:tasche} shows the three distributions from Example~\ref{ex:hom} for the case $n = 100$. 
It is no surprise that compared to the unconditional distribution the masses of the other two distributions are
significantly shifted to the right. But the `stressed input PDs' distribution seems to have heavier tails
than the `two defaults scenario' distribution. 
\begin{table}
\begin{center}
\caption{\small{}Characteristics of unconditional, conditional on two defaults and 
`stressed input PDs' distributions of numbers of defaults in portfolios of $n =10$, $n=100$ and $n=1000$ 
obligors. All obligors have
unconditional PD~1\%. The model is one-factor CreditRisk$^+$ with factor standard deviation 0.8.}\label{tab:1}
\vspace{1ex}
\begin{tabular}{|l||r|r|r|}
\hline
$n = 10$ & Unconditional & `Two defaults scenario' & `Stressed input PDs' \\ \hline
Probability of no default &  0.9076 & 0.8017 &   0.8083 \\ \hline
Mean  &  0.1000 & 0.2280  &  0.2280 \\ \hline
Standard deviation &     0.3262&  0.4925  &  0.5111 \\ \hline
99\%-quantile & 1 & 2  &  2 \\ \hline\hline
$n = 100$ & Unconditional & `Two defaults scenario' & `Stressed input PDs' \\ \hline
Probability of no default  &   0.4616 & 0.1716  &  0.2451 \\ \hline
Mean  &  1.0000  & 2.2800  &  2.2800 \\ \hline
Standard deviation &     1.2806 & 1.9337 &   2.3679\\ \hline
99\%-quantile & 5 & 8 &   10 \\ \hline\hline
$n = 1000$  & Unconditional & `Two defaults scenario' & `Stressed input PDs' \\ \hline
Probability of no default &    0.0438 & 0.0008 &   0.0137\\ \hline
Mean  &  10.0000   & 22.8000 &  22.8000\\ \hline
Standard deviation  &     8.6023 & 12.9892 &  18.8546\\ \hline
99\%-quantile & 39 & 63 &   87\\ \hline
\end{tabular}
\end{center} 
\end{table}

Table~\ref{tab:1} affirms this observation. The results from the table suggest that the difference between the
`two defaults scenario' and `stressed input PDs' distributions increases with growing portfolio size. The 
`stressed input PDs' distribution becomes markedly more widespread and heavier-tailed than the `two defaults scenario'
distribution. A possible explanation could be that the `two defaults scenario'  more appropriately takes account of diversification effects in the higher order
joint probabilities of default that strongly impact the tail of the distribution because 
`two defaults' is constructed as a proper 
conditional distribution.

With the following example, we study a more heterogenous portfolio, with a range of different PDs, different 
exposures and dependence created by two economic factors. We choose standard deviations of 0.4 and 1.2 respectively for
the two factors. According to \citet{Merino&Nyfeler2004}, this choice reflects the lower and upper bounds  
of the range of observable default rate volatilities.
\begin{example}\label{ex:het}
We assume the setting of Theorem~\ref{tas_th_3}, this time with the following specifics:
\begin{itemize}
\item There are $60$ obligors $B_1, \ldots, B_{60}$ all with PD~$p_i = 2.5\%$, $30$ obligors $B_{61}, \ldots, B_{90}$
all with PD~$p_i=1\%$ and $10$ obligors $B_{91}, \ldots, B_{100}$ all with PD~$p_i=0.5\%$. 
There are two economic factors $S_1$ and $S_2$ such that
the conditional Poisson distribution of the default indicator $D_{B_i}$ is given by the intensity
$p^S_{B_i} = p_i\,(w_i\,S_1+(1-w_i)\,S_2)$ with $w_i=0.75$ for all $i = 1, \ldots, 100$.
\item Two further obligors $A_1$ and $A_2$  are known to have 
defaulted. We assume they had PDs $q_i = 1\%$, $i=1,2$ and that their default intensities were given by
$q_{A_i}^S = q_i\,(v_i\,S_1+(1-v_i)\,S_2)$. We assume $v_1=v_2$. Results are calculated for two different values of
$v_1$, namely $v_1=0.75$ and $v_1=0.25$. 
\item The factor $S_1$ is Gamma-distributed with parameters $(1/1.2^2, 1.2^2) = (\alpha_1, 1/\alpha_1)$, 
$S_2$ is Gamma-distributed with parameters $(1/0.4^2, 0.4^2) = (\alpha_2, 1/\alpha_2)$.
\end{itemize}
\end{example}
Due to the heterogeneity of the portfolio in Example~\ref{ex:het}, it is not possible to represent the 
loss distribution of the loss variable $X$ in closed form. In order to calculate the unconditional, conditional
on the two defaults and `stressed probabilities of default' distributions of $X$ as in Example~\ref{ex:hom}, therefore
we take recourse to numerically inverting the respective characteristic functions by 
Fast Fourier Transform (FFT)\footnote{%
Alternatively, we could have made use of refined versions of the Panjer algorithm as described in
\citet{haaf2004numerically} or \citet[][Section~5.5]{gerhold2010generalization}.}. 
In all three cases, the shape of the characteristic function of the distribution is given by \eqref{tas_eq_gen} 
with $z = e^{i\,t}$, $t\in \mathbb{R}$. The algorithm we apply for the calculations is described in Section~4.7 of
\citet{Rolski&al}. The moderate size of the portfolio and the relatively small total exposure of the portfolio 
allow us to choose the total exposure plus 1 as the truncation point for the discrete Fourier transform. 
Indeed, the probabilities of the high losses close to the total exposure are so small that there is no
need for any refinements of the algorithm to control the aliasing error \citep[][Section~2.2]{embrechts2009panjer}.
\begin{table}
\begin{center}
\caption{\small{}Characteristics of unconditional, conditional on two defaults and 
`stressed input PDs' loss distributions of the portfolio described in 
Example~\ref{ex:het}.  The model is two-factors CreditRisk$^+$.}\label{tab:2}
\vspace{1ex}
\begin{tabular}{|l||r|r|r|}
\hline
& \multicolumn{3}{|c|}{Weak dependence of defaults and portfolio} \\ \hline\hline
 & Unconditional & `Two defaults scenario' & `Stressed input PDs' \\ \hline
Probability of no default &  0.2986 & 0.1769 &   0.1731 \\ \hline
Mean  &  4.0000 & 6.7173  &  6.7173 \\ \hline
Standard deviation &     6.4900 &  9.0172  &  9.2574 \\ \hline
99\%-quantile & 30 & 41  &  43 \\ \hline\hline
 & \multicolumn{3}{|c|}{Strong dependence of defaults and portfolio} \\ \hline\hline
 & Unconditional & `Two defaults scenario' & `Stressed input PDs' \\ \hline
Probability of no default &  0.2986 & 0.0545 &   0.0801 \\ \hline
Mean  &  4.0000 & 11.4514  &  11.4514 \\ \hline
Standard deviation &     6.4900 &  11.5349  &  13.8041 \\ \hline
99\%-quantile & 30 & 52  &  64 \\ \hline
\end{tabular}
\end{center} 
\end{table}

Table~\ref{tab:2} shows the results of the calculations for Example~\ref{ex:het}. Results are reported for
two different scenarios of dependence between the defaults and the rest of the portfolio:
\begin{itemize}
\item `Weak dependence of defaults and portfolio' scenario. By construction, the obligors $B_i$ in the portfolio
depend stronger on the economic factor $S_1$ (weight 0.75) than on the factor $S_2$ (weight 0.25). In the 
`weak dependence' scenario, the defaulters $A_1$ and $A_2$ depend weakly on $S_1$ (weight 0.25) and stronger
on $S_2$ (weight 0.75).
\item `Strong dependence of defaults and portfolio' scenario. Here the defaulters have the same dependence
on the economic factors as the obligors in the portfolio.
\end{itemize}
In both dependence scenarios, the impact of conditioning on defaults on the tails of the loss distributions is
strong but it is much stronger in the case of strong dependence. In the weak dependence scenario the
shapes of the conditional `two defaults scenario' loss distribution and the `stressed input PDs' 
distribution seem to be almost
equal. In contrast, in the strong dependence scenario the tail of the `stressed input PDs' distribution appears to
be much heavier than the tail of the `two defaults scenario' distribution. Note that as stated in
Proposition~\ref{pr:EqualMeans} in both Table~\ref{tab:1} and Table~\ref{tab:2} the means of the 
`two defaults scenario' and the `stressed input PDs' distributions always are equal.


\section{Conclusions}\label{sec:concl}

We have studied the way in which defaults impact a credit portfolio loss distribution in the CreditRisk$^+$ framework,
by looking at the loss distribution conditional on some -- one or two in this paper -- of defaults.
While the derived formulas are not necessarily easy to implement, they provide nonetheless insight 
into the details of how the default scenarios impact the conditional portfolio loss distribution.

The results of this paper can be used for specific stress scenario analyses that are intended to identify
whether large credit exposures besides having an obvious size impact additionally contribute to sector risk
concentrations. Another more indirect application of the results would be to use them to check the accuracy
of alternative approaches to such default scenario analyses. One potential alternative approach is Monte Carlo
portfolio simulation which would suffer from rare event effects when deployed for estimating loss distributions
conditional on two or more defaults. 

Another alternative could be to calculate for each obligor the
probability of default conditional on the joint default of a fixed set of obligors and then to use these
conditional probabilities of default as input parameters to a portfolio model. This ``stressed probabilities of default''
approach is unbiased but ignores the exact dependence between the default events of the obligors considered 
defaulted under the scenario and 
the economic factors commonly used for modeling dependence in credit portfolio models. Therefore, the approach is
principally inaccurate. Numerical examples show that the inaccuracy may be significant but tends to overestimate
tail losses and hence to err on the conservative side.



\end{document}